\documentclass
[showpacs,superscriptaddress,prl,twocolumn,tightenlines,balancelastpage,10pt,a4paper]{revtex4}
\usepackage{amsmath}
\usepackage{amsfonts}
\usepackage{amssymb}
\usepackage{graphicx}%

\begin{document}
\title{Experimental Violation of Bell Inequality beyond Cirel'son's Bound}
\author{Yu-Ao Chen}
\email{yuao@physi.uni-heidelberg.de}
\affiliation{Physikalisches
Institut, Universit\"{a}t Heidelberg, Philosophenweg 12, D-69120
Heidelberg, Germany}
\author{Tao Yang}
\affiliation{Physikalisches Institut, Universit\"{a}t Heidelberg,
Philosophenweg 12, D-69120 Heidelberg, Germany} \affiliation{Hefei
National Laboratory for Physical Sciences at Microscale,
Department of Modern Physics, University of Science and Technology
of China, Hefei, 230027, People's Republic of China}
\author{An-Ning Zhang}
\affiliation{Hefei National Laboratory for Physical Sciences at
Microscale, Department of Modern Physics, University of Science
and Technology of China, Hefei, 230027, People's Republic of
China}
\author{Zhi Zhao}
\affiliation{Hefei National Laboratory for Physical Sciences at
Microscale, Department of Modern Physics, University of Science
and Technology of China, Hefei, 230027, People's Republic of
China}
\author{Ad\'{a}n Cabello}
\email{adan@us.es} \affiliation{Departamento de F\'{\i}sica Aplicada
II, Universidad de Sevilla, E-41012 Sevilla, Spain}
\author{Jian-Wei Pan}
\affiliation{Physikalisches Institut, Universit\"{a}t Heidelberg,
Philosophenweg 12, D-69120 Heidelberg, Germany} \affiliation{Hefei
National Laboratory for Physical Sciences at Microscale,
Department of Modern Physics, University of Science and Technology
of China, Hefei, 230027, People's Republic of China}
\date{\today}


\begin{abstract}
The correlations between two qubits belonging to a three-qubit
system can violate the Clauser-Horne-Shimony-Holt-Bell inequality
beyond Cirel'son's bound [A. Cabello, Phys. Rev. Lett. {\bf 88},
060403 (2002)]. We experimentally demonstrate such a violation by 7
standard deviations by using a three-photon polarization-entangled
Greenberger-Horne-Zeilinger state produced by Type-II spontaneous
parametric down-conversion. In addition, using part of our results,
we obtain a violation of the Mermin inequality by 39 standard
deviations.
\end{abstract}


\maketitle




As stressed by Peres \cite{Peres93}, Bell inequalities
\cite{Bell64,Bell87} have nothing to do with quantum mechanics. They
are constraints imposed by local realistic theories on the values of
linear combinations of the averages (or probabilities) of the
results of experiments on two or more separated systems. Therefore,
when examining data obtained in experiments to test Bell
inequalities, it is legitimate to do it from the perspective (i.e.,
under the assumptions) of local realistic theories, without any
reference to quantum mechanics. This approach leads to some
apparently paradoxical results. A remarkable one is that, while it
is a standard result in quantum mechanics that no quantum state can
violate the Clauser-Horne-Shimony-Holt (CHSH) Bell inequality
\cite{CHSH69} beyond Cirel'son's bound, namely $2 \sqrt 2$
\cite{Cirelson80}, the correlations between two qubits belonging to
a three-qubit system can violate the CHSH-Bell inequality beyond $2
\sqrt{2}$ \cite{Cabello02}. In particular, if we use a three-qubit
Greenberger-Horne-Zeilinger (GHZ) state \cite{GHZ89}, we can even
obtain the maximum allowed violation of the CHSH-Bell inequality,
namely $4$ \cite{Cabello02}.

In this Letter, we report the first observation of a violation of
the CHSH-Bell inequality beyond Cirel'son's bound by using a
three-photon polarization-entangled GHZ state produced by Type-II
spontaneous parametric down-conversion. In addition, since the
experiment also provides all the data required for testing Mermin's
three-party Bell inequality \cite{Mermin90}, we use our results to
demonstrate the violation of this inequality.

The main idea behind the CHSH-Bell inequality \cite{CHSH69} is that,
in local realistic theories, the absolute value of a particular
combination of correlations between two distant particles $i$ and
$j$ is bounded by 2:
\begin{align}
\vert C\left( A,B\right) -mC\left( A,b\right) & -nC\left( a,B\right)
\nonumber\\
& -mnC\left( a,b\right) \vert\leqslant2\label{CHSH1}%
\end{align}
where \emph{m} and \emph{n} can be either $-1$ or $1$, and $A$ and
$a$ ($B$ and $b$) are physical observables taking values $-1$ or
$1$, referring to local experiments on particle $i$ ($j$). The
correlation $C\left( A,B\right)$ of $A$ and $B$ is defined as
\begin{align}
C\left( A,B\right) & =P_{AB}\left( 1,1\right) -P_{AB}\left( 1,-1\right)
\nonumber\\
& -P_{AB}\left( -1,1\right) +P_{AB}\left( -1,-1\right),
\label{correlation}
\end{align}
where $P_{AB}\left( 1,-1\right)$ denotes the joint probability of
obtaining $A=1$ and $B=-1$ when $A$ and $B$ are measured.

Cirel'son proved that, for a two particle system prepared in any
quantum state, the absolute value of the combination of quantum
correlations appearing in the inequality (\ref{CHSH1}) is bounded
by 2$\sqrt{2}$ \cite{Cirelson80}. However, assuming local
realistic theories' point of view, the correlations predicted by
quantum mechanics between two distant qubits belonging to a
three-qubit system can violate the CHSH-Bell inequality beyond
Cirel'son's bound \cite{Cabello02}.

In our experiment, the three distant qubits are
polarization-entangled photons prepared in the GHZ state:
\begin{equation}
\vert\Psi\rangle=\frac{1}{\sqrt{2}}\left(
|H\rangle|H\rangle|H\rangle +|V\rangle|V\rangle|V\rangle\right),
\label{GHZ}
\end{equation}
where $H$ ($V$) denotes horizontal (vertical) linear polarization.
During the experiment, we will analyze the polarization of each
photon in one of two different basis: either in the $X$ basis,
which is defined as the linear
polarization basis $H/V$ rotated by $45^{0}$, which is denoted as
$H^{\prime}/V^{\prime}$; or in the $Y$ basis,
which is defined as the circular polarization basis $R/L$
(right-hand/left-hand). These polarization bases can be expressed in
terms of the $H/V$ basis as
\begin{align}
& \left\vert H^{\prime}\right\rangle =\frac{1}{\sqrt{2}}\left(|H\rangle
+|V\rangle\right) , \quad\left\vert V^{\prime}\right\rangle =\frac{1}{\sqrt
{2}}\left(|H\rangle-|V\rangle\right), \nonumber\\
& \left\vert R\right\rangle =\frac{1}{\sqrt{2}}\left(|H\rangle
+i|V\rangle\right) , \quad\left\vert L\right\rangle =\frac{1}{\sqrt{2}}\left(
|H\rangle-i|V\rangle\right) . \label{basis}%
\end{align}
The measurement results $H^{\prime}$ ($R$) and $V^{\prime}$ ($L$)
are denoted by $1$ and $-1$, respectively.


\begin{figure}[ptb]
\begin{center}
\includegraphics[
height=2.9in, width=3.277in
]{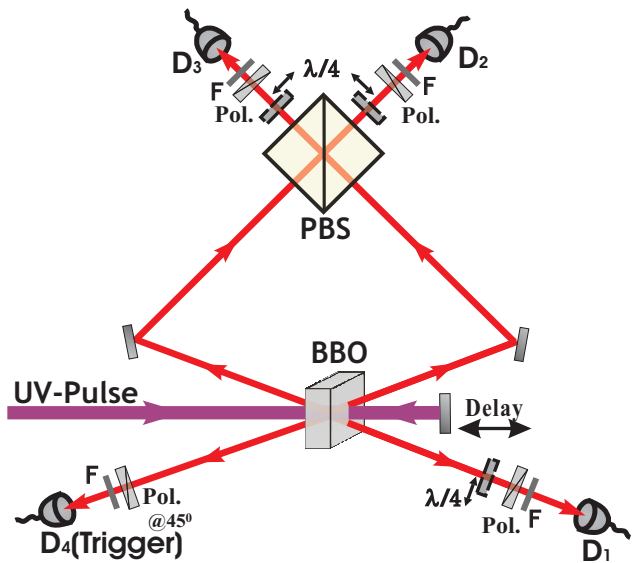}
\end{center}
\caption{Experimental setup for generating three-photon GHZ
states. An UV pulse passes twice through the BBO crystal to
generate two pairs of polarization-entangled photons by Type-II
spontaneous parametric down-conversion used to perform the
preparation of three-photon GHZ state. The UV laser with a central
wavelength of 394 nm has pulse duration of 200fs, a repetition
rate of 76 MHz, and an average pump power of 400 mW. We observe
about $2\times10^{4}$ entangled pairs per second behind $3.6$ nm
filters (F) of central wavelength $788$ nm. Polarizers (Pol.) and
quarter wave plates ($\lambda/4$) in front of the
detectors are used for performing $X$ or $Y$ measurement.}%
\label{setup}%
\end{figure}


To generate the three-photon GHZ state (\ref{GHZ}) we use the
technique developed in previous experiments
\cite{BPDWZ99,PDGWZ01}. The experimental setup for generating
three-photon entanglement is shown in Fig.~\ref{setup}. A pulse of
ultraviolet (UV) light passes through a beta-barium borate (BBO)
crystal twice to produce two polarization-entangled photon pairs,
where both pairs are in the state
\begin{equation}
\left\vert \Psi_2\right\rangle =1/\sqrt{2}\left( |H\rangle|H\rangle
+|V\rangle|V\rangle\right).
\end{equation}
One photon out of each pair is then steered to a polarization beam
splitter (PBS) where the path lengths of each photon have been
adjusted (by scanning the delay position) so that they arrive
simultaneously. After the two photons pass through the PBS, and exit
it by a different output port each, there is no way whatsoever to
distinguish from which emission each of the photons originated, then
correlations due to four-photon GHZ entanglement
\begin{equation}
\left\vert \Psi_4\right\rangle =1/\sqrt {2}\left(
|H\rangle|H\rangle|H\rangle|H\rangle+|V\rangle|V\rangle
|V\rangle|V\rangle\right)
\end{equation}
can be observed \cite{PDGWZ01}. After that, by performing a
$|H^{\prime}\rangle$ polarization projective measurement onto one
of the four outputs, the remaining three photons will be prepared
in the desired GHZ state (\ref{GHZ}).


\begin{figure}[ptb]
\begin{center}
\includegraphics[width=3.6in ]{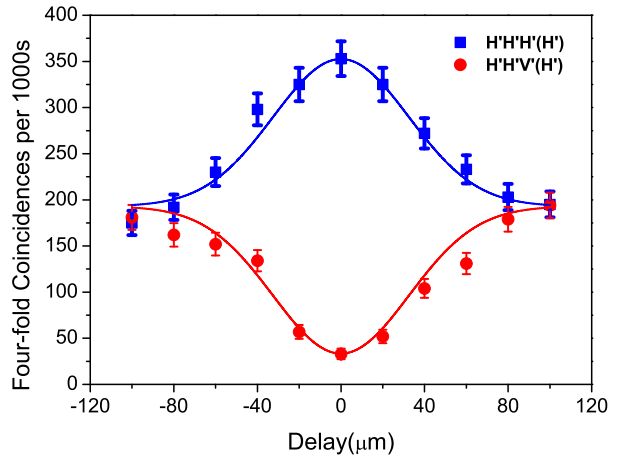}
\end{center}
\caption{Typical experimental results for polarization measurements
on all three photons in a $X$ basis triggered by the photon $4$ at
the $H'$ polarization. The coincidence rates of $H'H'H'$ and
$H'H'V'$ components are shown as a function of the pump delay mirror
position. The high visibility obtained at zero delay implies that
three photons are indeed in a coherent superposition.}
\label{coherent}
\end{figure}


In the experiment, the observed fourfold coincident rate of the
desired component $HHHH$ or $VVVV$ is about $1.4$ per second. By
performing a $H'$ projective measurement at photon $4$ as the
trigger of the fourfold coincident, the ratio between any of the
desired events $HHH$ and $VVV$ to any of the 6 other nondesired
ones, e.g., $HHV$, is about $65:1$. To confirm that these two events
are indeed in a coherent superposition, we have performed
polarization measurements in $X$ basis. In Fig.~\ref{coherent}, we
compare the count rates of $H'H'H'$ and $H'H'V'$ components as we
move the delay mirror (Delay) by the trigger photon $4$ at the $H'$
polarization. The latter component is suppressed with a visibility
of $83\%$ at zero delay, which confirms the coherent superposition
of $HHH$ and $VVV$.


For each three-photon system prepared in the state (\ref{GHZ}), we
will define as photons $i$ and $j$ those two giving the result $-1$
when making $X$ measurement on all three photons; the third photon
will be denoted as $k$. If all three photons give the result $1$,
photons $i$ and $j$ could be any pair of them. Since no other
combination of results is allowed for the state (\ref{GHZ}), $i$ and
$j$ are well defined for every three-photon system.

We are interested in the correlations between two observables, $A$
and $a$, of photon $i$ and two observables, $B$ and $b$, of photon
$j$. In particular, let us choose $A=X _{i}$, $a=Y _{i}$, $B=X
_{j}$, and $b=Y _{j}$, where $X _{q}$ and $Y _{q}$ are the
polarizations of photon $q$ along the basis $X$ and $Y$,
respectively. The particular CHSH-Bell inequality (\ref{CHSH1}) we
are interested in is the one in which $m=n=y _{k}$, where $y _{k}$
is one of the possible results, $-1$ or $1$ (although we do not know
which one), of measuring $Y_{k}$. With this choice we obtain the
CHSH-Bell inequality
\begin{align}
\vert C & \left( X_{i},X_{j}\right) -y_{k} C\left(
X_{i},Y_{j}\right)
\nonumber\\
& -y_{k} C\left( Y_{i},X_{j}\right) -C\left( Y_{i},Y_{j}\right)
\vert \leqslant2, \label{CHSH3}
\end{align}
which holds for local realistic theories, regardless of the
particular value, either $-1$ or $1$, of $y_{k}$.

We could force photons $i$ and $j$ to be those in locations $1$ and
$2$, by measuring $X$ on the photon in location $3$, and then
selecting only those events in which the result of this measurement
is $1$. This procedure guarantees that our definition of photons $i$
and $j$ is physically meaningful. However, it does not allow us to
measure $Y$ on photon $k$. The key point for testing inequality
(\ref{CHSH3}) is noticing that we do not need to know in which
locations are photons $i$, $j$, and $k$ for every three-photon
system. We can obtain the required data by performing suitable
combinations of measurements of $X$ or $Y$ on the three photons. In
order to see this, let us first translate inequality (\ref{CHSH3})
into the language of joint probabilities. Assuming that the expected
value of any local observable cannot be affected by anything done to
a distant particle, the CHSH-Bell inequality (\ref{CHSH3}) can be
transformed into a more convenient experimental inequality
\cite{CH74,Mermin95}:
\begin{align}
-1\leqslant & P_{X_{i}X_{j}}(-1,-1)-P_{X_{i}Y_{j}}(-1,-y_{k})\nonumber\\
& -P_{Y_{i}X_{j}}(-y_{k},-1)-P_{Y_{i}Y_{j}}(y_{k},y_{k})\leqslant
0.
\label{CHSH4}
\end{align}
The bounds $l$ of inequalities (\ref{CHSH1}) and (\ref{CHSH3}) are
transformed into the bounds $(l-2)/4$ of inequality (\ref{CHSH4}).
Therefore, the local realistic bound in (\ref{CHSH4}) is $0$,
Cirel'son's bound is $(\sqrt{2}-1)/2$, and the maximum value is
$1/2$.

To measure the inequality (\ref{CHSH4}), we must relate the four
joint probabilities appearing in (\ref{CHSH4}) to the probabilities
of coincidences in a experiment with three spatial locations, $1$,
$2$, and $3$. For instance, it can be easily seen that
\begin{align}
P_{X_{i}} & _{X_{j}}(-1,-1)=\nonumber\\
& P_{X_{1}X_{2}X_{3}}(1,-1,-1)+P_{X_{1}X_{2}X_{3}}(-1,1,-1)\nonumber\\
& +P_{X_{1}X_{2}X_{3}}(-1,-1,1)+P_{X_{1}X_{2}X_{3}}(-1,-1,-1).
\label{prob1}
\end{align}
In addition, $P_{X_{i}Y_{j}}(-1,-y_{k})$ and
$P_{Y_{i}X_{j}}(-y_{k},-1)$ are both less than or equal to
\begin{align}
P_{X_{1} Y_{2}} & _{Y_{3}}(-1,1,-1)+P_{X_{1} Y_{2} Y_{3}}(-1,-1,1)\nonumber\\
& +P_{Y_{1} X_{2} Y_{3}}(1,-1,-1)+P_{Y_{1} X_{2} Y_{3}}(-1,-1,1)\nonumber\\
& +P_{Y_{1} Y_{2} X_{3}}(1,-1,-1)+P_{Y_{1} Y_{2} X_{3}}(-1,1,-1).
\label{prob2}
\end{align}
Finally,
\begin{align}
P_{Y_{i}} & _{Y_{j}}(y_{k},y_{k})=P_{Y_{1} Y_{2}
Y_{3}}(1,1,1)+P_{Y_{1} Y_{2} Y_{3}}(-1,-1,-1).
\label{prob3}
\end{align}
Therefore, by performing measurements in $5$ specific configurations
($XXX$, $XYY$, $XYX$, $YXX$, and $YYY$), we can obtain the value of
the middle side of inequality (\ref{CHSH4}).

In the state ({\ref{GHZ}}), the first three probabilities in the
right-hand of (\ref{prob1}) are expected to be $1/4$, and the fourth
is expected to be zero; the six probabilities in (\ref{prob2}) are
expected to be zero, and the two probabilities in the right hand
side of (\ref{prob3}) are expected to be $1/8$. Therefore, the
middle side of inequality (\ref{CHSH4}) is expected to be $1/2$,
which means that the left-hand side of inequality (\ref{CHSH3}) is
$4$, which is not only beyond Cirel'son's bound, $2 \sqrt{2}$, but
is also the maximum possible violation of inequality (\ref{CHSH3}).


\begin{figure}
[ptb]
\begin{center}
\includegraphics[trim=0.994243in 0.198588in 0.449956in 0.501064in,
width=2.8in]{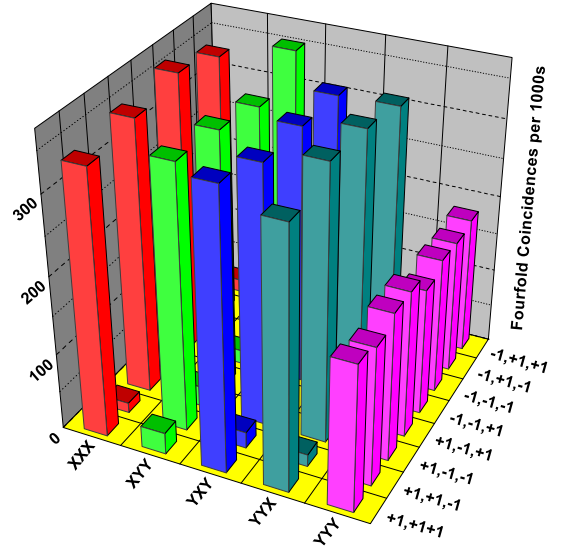}
\end{center}
\caption{Experimental results observed for the 5 required
configurations: $XXX$, $XYY$ $YXY$, $YYX$, and $YYY$.}
\label{result}
\end{figure}


The experiments consists of performing measurements in $5$ specific
configurations. As shown in Fig.~\ref{setup}, we use polarizers
oriented at $\pm45^{0}$ and $\lambda/4$ plates to perform $X$ or $Y$
measurements. For these $5$ configurations, the experimental results
for all possible outcomes are shown in Fig.~\ref{result}.

Substituting the experimental results (shown in Fig.~\ref{result})
into the right-hand side of (\ref{prob1}), we obtain
\begin{align}
P_{X_{i}} & _{X_{j}}(-1,-1)=0.738\pm0.012.
\label{ineq1}
\end{align}
Similarly, substituting the experimental results in (\ref{prob2}),
we obtain
\begin{align}
P_{X_{i}Y_{j}}(-1,-y_{k})\leqslant0.072\pm0.007,\nonumber\\
P_{Y_{i}X_{j}}(-y_{k},-1)\leqslant0.072\pm0.007.
\label{ineq2}
\end{align}
Finally, substituting the experimental results in (\ref{prob3}), we
obtain
\begin{align}
P_{Y_{i}Y_{j}}(y_{k},y_{k})=0.254\pm0.011.
\label{ineq3}
\end{align}
Therefore, the prediction for the middle side of (\ref{CHSH4}) is
greater than or equal to $0.340\pm0.019$, and the prediction for the
right-hand side of (\ref{CHSH3}) is greater than or equal to
$3.36\pm0.08$, which clearly violates Cirel'son's bound by $7$
standard deviations.

In addition, using part of the results contained in
Fig.~\ref{result}, we can test the three-particle Bell inequality
derived by Mermin \cite{Mermin90},
\begin{align}
\vert C & \left(X_1,Y_2,Y_3\right)+C\left(Y_1,X_2,Y_3\right)
\nonumber\\
& +C\left(Y_1,Y_2,X_3\right)-C\left(X_1,X_2,X_3\right) \vert
\leqslant2. \label{Mermin}
\end{align}
From the results in Fig.~\ref{result} we obtain $3.57\pm0.04$ for
the left-hand side of (\ref{Mermin}), which is a violation of
inequality (\ref{Mermin}) by $39$ standard deviations. Note that the
experiment for observing the violation beyond Cirel'son's bound also
requires performing measurements in an additional configuration
($YYY$).

In conclusion, we have demonstrated a violation of the CHSH-Bell
inequality beyond Cirel'son's bound. It should be emphasized that
such a violation is predicted by quantum mechanics and appears
when examining the data from the perspective of local realistic
theories \cite{Cabello02}. In addition, it should be stressed that
the reported experiment is different as previous experiments to
test local realism involving three or four-qubit GHZ states
\cite{PBDWZ00,ZYCZZP03}, since it is based on a definition of
pairs which is conditioned to the result of a measurement on a
third qubit, and requires performing measurements in additional
configurations.




This work was supported by the Alexander von Humboldt Foundation,
the Marie Curie Excellence Grant from the EU, the Deutsche Telekom
Stiftung, the NSFC, and the CAS. A.C. acknowledges additional
support from Projects No.~FIS2005-07689 and No.~FQM-239.


\end{document}